\documentclass[12pt]{article} % Para Notas de F\'isica 
\usepackage[esperanto,portuguese]{babel}
\usepackage{epsfig}
\usepackage{icomma}
\newcommand{\ppunu}[1]{#1} \newcommand{\ppdu}[1]{}
\ppdu{\markright{CBPF-NF-023/09}}

\newcommand{\comentario}[1]{}
\newcommand{\Sch}{Schwarz\-schild}
\newcommand{\dd}{\mathrm{d}}

\ppdu{
\newlength{\pplw}\setlength{\pplw}{0.453\textwidth}
\newlength{\pprw}\setlength{\pprw}{0.507\textwidth}

\newcommand{\ppp}{\ParallelPar}
\newcommand{\ppn}{\noindent}              %p/ nao fazer parag
\newcommand{\ppl}[1]{\ParallelLText{\selectlanguage{esperanto}#1}}
\newcommand{\ppr}[1]{\ParallelRText{\selectlanguage{portuguese}#1}\ppp}
\newcommand{\ppln}[1]
{\ParallelLText{\ppn \selectlanguage{esperanto}#1}} %p/ nao parag
\newcommand{\pprn}[1]
{\ParallelRText{\ppn \selectlanguage{portuguese}#1}\ppp} %p/ nao parag

\newcommand{\ppR}[1]{\ParallelRText{#1}}

\newcommand{\ppsection}[3][0ex]{\vspace{2em} 
\ppl{\section{#2} \vspace{#1}} \ppa \nopagebreak
\ppR{\section{#3}} \ppp \nopagebreak}

\newcommand{\bea}{\vspace{-1ex}\begin{eqnarray}}
\newcommand{\eea}{\end{eqnarray}}
}

\ppunu{
\newcommand{\ppl}[1]{\selectlanguage{esperanto}#1}
\newcommand{\ppln}[1]{\noindent \selectlanguage{esperanto}#1}
\newcommand{\ppr}[1]{\selectlanguage{portuguese}}
\newcommand{\pprn}[1]{\noindent \selectlanguage{portuguese}}
\newcommand{\ppsection}[3][0ex]{\section{#2}}

\newcommand{\bea}{\begin{eqnarray}}
\newcommand{\eea}{\end{eqnarray}}
}

\title{{\bf Doppleraj efikoj \^ce \Sch \ppdu{\\ Efeitos Doppler em \Sch}}}
\author{F.M. Paiva \\ 
{\small Departamento de F\'isica, Unidade Humait\'a II, Col\'egio Pedro II} \\
{\small Rua Humait\'a 80, 22261-040  Rio de Janeiro-RJ, Brasil; fmpaiva@cbpf.br} 
\vspace{.7ex} \\
A.F.F. Teixeira \\
{\small Centro Brasileiro de Pesquisas F\'isicas} \\
{\small 22290-180 Rio de Janeiro-RJ, Brasil; teixeira@cbpf.br}}
\begin{document}
\selectlanguage{esperanto}
\maketitle
\thispagestyle{empty}

\begin{abstract}\selectlanguage{esperanto}
Movado de korpoj kaj lumo estas studitaj en gravita kampo de \Sch. Pluraj Doppleraj efikoj estas priskribitaj.

\ppdu{\selectlanguage{portuguese}
Movimentos de corpos e raios luminosos s\~ao estudados no campo gravitacional de \Sch. V\'arios efeitos Doppler s\~ao descritos.}
%\ppunu{Moving bodies and light rays are studied in the gravitational field of \Sch. Several Doppler effects are described.}
\end{abstract}

%\selectlanguage{portuguese}

\ppdu{
\begin{Parallel}[v]{\pplw}{\pprw}
%\begin{Parallel}[v]{}{}
}

%\ppl{Blo -- blo -- blo.} 
%\ppr{} 

\ppdu{\section*{\vspace{-2em}}\vspace{-2ex}}   %PORQUE PRECISO DISTO ?

\ppsection[0.6ex]{Enkonduko}{Introdu\c c\~ao}

\ppln{En~\cite{reltemp1}--\cite{lumaebeno} ni studis la tempon ^ce la special-relativeco. Alia artikolo~\cite{reltemp2} nia estas fini^ganta, zorgante pri difinojn de pluraj malsamaj intertempoj ^ce la ^general-relativeco. Ni nun prezentas specifigon de~\cite{reltemp2} por spacotempo de \Sch,} 
\pprn{Em~\cite{reltemp1}--\cite{lumaebeno} n\'os estudamos o tempo na relatividade especial. Um outro artigo~\cite{reltemp2} est\'a em conclus\~ao, tratando de defini\c c\~oes de v\'arios diferentes intertempos na relatividade geral. N\'os apresentamos agora uma especializa\c c\~ao do~\cite{reltemp2} para o espa\c cotempo de \Sch,}

\bea                                                      \label{Schw}%01 
\dd s^2 = \left(1-\frac{\rho}{r}\right)(c\,\dd t)^2 - \left(1-\frac{\rho}{r}\right)^{-1}(\dd r)^2 - r^2(\dd\theta)^2 - r^2{\sin}^2\theta\,(\dd\varphi)^2\,, 
\hspace{3mm} \rho:=\frac{2\,Gm}{c^2}\,. 
\eea 

\ppln{Ni zorgos pri pluraj movadoj en kampo (\ref{Schw}), en ebeno $\theta=\pi/2$ de regiono $r>\rho$.} 
\pprn{N\'os vamos tratar de v\'arios movimentos no campo (\ref{Schw}), no plano $\theta=\pi/2$ da regi\~ao $r>\rho$.}

\ppl{Memoru ke movado de materio havas $\dd s^2$ pozitivan kaj tiuokaze oni difinas infiniteziman propran intertempon $\dd\tau$ kiel $\dd s=c\,\dd\tau$; alie, movado de lumo havas $\dd s=0$.} 
\ppr{Relembre que um movimento de mat\'eria tem $\dd s^2$ positivo, e neste caso define-se um intervalo infinitesimal de tempo pr\'oprio $\dd\tau$ como $\dd s=c\,\dd\tau$; por outro lado, um  movimento de luz tem $\dd s=0$.}

\ppl{En sekcio 2 ni priskribas konstruadon de koordinat\-horlo^goj de metriko (\ref{Schw}). En sekcio 3 ni difinas Doppleran efikon kaj prezentas la nemikse gravitan Doppleran faktoron en tiu metriko. En sekcio 4 ni skribas la diferencialajn ekvaciojn de geodezoj, kaj emfazas la gravecon de konstantoj $E$ kaj $h$ de integro. En sekcio 5 ni studas cirklajn geodezojn, kaj sekcio 6 studas aliajn cirklajn movadojn. Sekcio 7 montras, ke propratempo de cirkle movi^ganta korpo fluas malpli rapide ol tio de korpo restanta en la sama radiuso. ^Gi priskribas anka^u Doppleran efikon inter du korpoj cirkle movi^gantaj en la sama ebeno, kaj kun la sama rotacirapido. Sekcio~8 pririgardas interesan specialan okazon de lasta sistemo, prezentante nulan Doppleran efikon. En sekcio 9 ni komparas la fluon de propratempo de radiuse geodeze movi^ganta korpo kun tio de ripozanta korpo, kaj sekcio 10 kalkulas Doppleran faktoron en tiu movado. Fine,  sekcio 11 aplikas niajn rezultojn al movado de satelitoj orbitante Teron.}
\ppr{Na se\c c\~ao 2 n\'os descrevemos a constru\c c\~ao de rel\'ogios de coordenada da m\'etrica (\ref{Schw}). Na se\c c\~ao 3 n\'os definimos efeito Doppler e apresentamos o fator Doppler puramente gravitacional nessa m\'etrica. Na se\c c\~ao 4 n\'os escrevemos as equa\c c\~oes diferenciais das geod\'esicas e enfatizamos a relev\^ancia das constantes $E$ e $h$ de integra\c c\~ao. Na se\c c\~ao 5 n\'os estudamos geod\'esicas circulares, e a se\c c\~ao 6  estuda outros movimentos circulares. A se\c c\~ao 7 mostra que o tempo pr\'oprio de um corpo em movimento circular flui menos r\'apido que aquele de um corpo parado no mesmo raio. Ela descreve tamb\'em o efeito Doppler entre dois corpos movendo-se circularmente no mesmo plano, e com mesma velocidade angular. A se\c c\~ao~8 examina um interessante caso especial deste \'ultimo sistema, apresentando efeito Doppler nulo. Na se\c c\~ao 9 n\'os comparamos o fluir do tempo pr\'oprio de um corpo em movimento geod\'etico radial com o de um corpo parado, e a se\c c\~ao 10 calcula o fator Doppler nesse movimento. Finalmente, a se\c c\~ao 11 aplica nossos resultados ao movimento de sat\'elites orbitando a Terra.}

\ppsection[0.6ex]{Koordinathorlo\^goj}{Rel\'ogios de coordenada}

\ppln{\^Car la koeficiento $g_{00}$ de metriko (\ref{Schw}) pendas de $r$, tial du koordinathorlo^goj de (\ref{Schw}) estas identikaj nur se ili estas en sama radiusa pozicio. Krome nur la koordinathorlo^goj de infinito  (kie $g_{00}=1$) estas normhorlo^goj. Se normhorlo^go (kiu montras $d\tau$) estas fiksata ^ce koordinathorlo^go (kiu montras $dt$) en finhava radiusa pozicio, d$\tau=\sqrt{g_{00}}\,\dd t$; ^car $g_{00}<1$, tial normhorlo^go mar^sas malpli rapide ol loka koordinathorlo^go.}
\pprn{Como o coeficiente $g_{00}$ da m\'etrica (\ref{Schw}) depende de $r$, dois rel\'ogios de coordenada de (\ref{Schw}) s\~ao id\^enticos somente se eles estiverem na mesma posi\c c\~ao radial. Al\'em disso, somente os rel\'ogios de coordenada do infinito (onde $g_{00}=1$) s\~ao rel\'ogios padr\~ao. Se um rel\'ogio padr\~ao (que mostra d$\tau$) for fixado ao lado de um rel\'ogio de coordenada (que mostra d$t$) em posi\c c\~ao radial finita, ent\~ao d$\tau=\sqrt{g_{00}}\,\dd t$\,; como $g_{00}<1$\,, o rel\'ogio padr\~ao tem andamento menos r\'apido que o rel\'ogio de coordenada local.}

\ppl{Demando: kiel estas konstruitaj la koordinathorlo^goj de metriko (\ref{Schw})? Respondo: (formala) ebleco estas unue fiksi unu {\it normhor\-lo\-^gon} en ^ciu punkto $[r,\theta,\varphi]$ de spaca teksa^jo; poste, uzante la rilaton $\dd\tau= \sqrt{g_{00}(r)}\,\dd t$, rapidigi la mar^son de ^ciu horlo^go la^u ^gia radiusa pozicio $r$. Alterne, la mekanismo de ^ciu horlo^go estas ne\^san\^gata, sed la ciferplato estas konvene ^san^gata por kompensi la fakton, ke $\dd t>\dd\tau.$ Fine sinkronigi ilin la^u Einstein~\cite{reltemp1}.}
\ppr{Uma pergunta: como s\~ao constru\'idos os rel\'ogios de coordenada da m\'etrica (\ref{Schw})? Res\-posta: uma possibilidade (formal) \'e primeiramente fixar um {\it rel\'ogio padr\~ao} em cada ponto $[r,\theta,\varphi]$ da trama espacial; depois, usando a rela\c c\~ao $\dd\tau= \sqrt{g_{00}(r)}\,\dd t$, apressar a marcha de cada rel\'ogio conforme sua posi\c c\~ao radial $r$. Alternativamente, o mecanismo de cada rel\'ogio n\~ao \'e mudado, mas a numera\c c\~ao no mostrador \'e apropriadamente mudada para compensar o fato que $\dd t>\dd\tau.$ Finalmente sincroniz\'a-los ao modo de Einstein~\cite{reltemp1}.}

\begin{figure}
\centerline{\epsfig{file=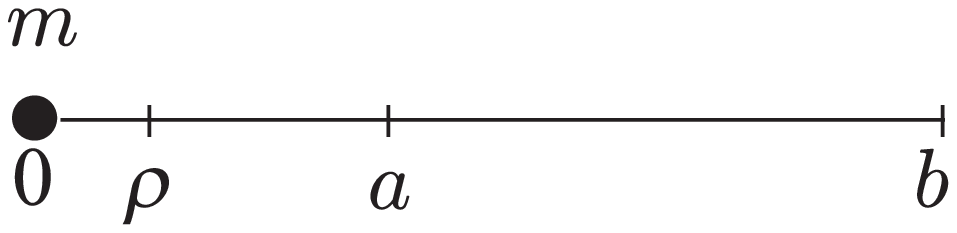,width=6cm}}
\selectlanguage{esperanto}\caption{\selectlanguage{esperanto}
Fonto restanta en pozicio $r=a$ eligas lumon en kampo de \Sch\ (\ref{Schw}); tiu lumo estas perceptata kun Dopplera faktoro (\ref{DopSchw}), per observanto restanta en pozicio $r=b$. 
\ppdu{\newline \selectlanguage{portuguese}Figura~\ref{FigDopEst}:
Uma fonte parada na posi\c c\~ao $r=a$ emite luz em um campo de \Sch\ (\ref{Schw}); essa luz \'e percebida com fator Doppler (\ref{DopSchw}), por um observador parado na posi\c c\~ao $r=b$.}}
\label{FigDopEst}
\end{figure}

\ppsection[0.6ex]{Nemikse gravita \ppdu{\\ }Dopplera efiko}{Efeito Doppler \\ puramente gravitacional}

\ppln{Frekvenco estas grava eco de luma ondo. Se frekvenco de lumo ricevata estas malsimila al tio de lumo eligita ni diras, ke okazas Dopplera efiko. Tiu efiko estas mezurata per la Dopplera faktoro, difinata kiel}
\pprn{A frequ\^encia \'e uma caracter\'istica importante de uma onda luminosa. Se a frequ\^encia de uma luz recebida for diferente daquela da luz emitida, n\'os dizemos que ocorre efeito Doppler. Esse efeito \'e medido pelo fator Doppler, definido como} 

\bea                                                       \label{Dnu}%02
D_{a\rightarrow b}:=\frac{\nu_{\,b}}{\nu_{a}}\,,
\eea 

\ppln{estante $\nu_a$ kaj $\nu_{\,b}$ la {\it propraj frekvencoj} de lumo igita el $a$ kaj observata en $b$, respektive. La efiko nomi^gas ru^gi^go se $D<1$, kaj violi^go se $D>1$.}
\pprn{sendo $\nu_a$ e $\nu_{\,b}$ as {\it frequ\^encias pr\'oprias} da luz emitida de $a$ e observada em $b$, respectivamente. O efeito se chama desvio para o vermelho se $D<1$, e desvio para o violeta se $D>1$.} 

\ppl{Simpla maniero por kalkuli $D$ estas fari, ke fonto eligu du sinsekvajn lumajn signalojn, apartitajn de infinitezima propra intertempo $\dd\tau_a$. Se tiuj signaloj estas ricevataj kun infinitezima propra intertempo $\dd\tau_b$\,, la Dopplera faktoro estas}
\ppr{Um modo simples para calcular $D$ \'e fazer com que a fonte emita dois sinais luminosos consecutivos, separados por um intertempo pr\'oprio infinitesimal $\dd\tau_a$. Se esses sinais forem recebidos com intertempo pr\'oprio infinitesimal $\dd\tau_b$\,, o fator Doppler \'e}

\bea                                                      \label{Dtau}%03
D_{a\rightarrow b}=\frac{\dd\tau_a}{\dd\tau_b}\,.
\eea

\ppl{Frekvenc^san^go pendas de interrapido de fonto-observanto, kaj anka^u de gravito. Speciale, se amba^u lumfonto kaj observanto restas, kaj gravito estas ne^san^ginta ($g_{\mu\nu}$ ne pendas de tempa koordinato), ni kalkulas la Doppleran faktoron tre facile. Fakte, tiuokaze la koordinata intertempo $\dd t$ inter la eligitaj signaloj kaj la ricevataj estas samaj \cite[pa\^go\,416]{Anderson}. ^Car por la du restantaj korpoj okazas $\dd\tau_a=\sqrt{g_{00}(a)}\,\dd t$ kaj $\dd\tau_b=\sqrt{g_{00}(b)}\,\dd t$\,, tial la (\ref{Dtau}) kaj (\ref{Schw}) faras (vidu  figuron~\ref{FigDopEst})}
\ppr{A mudan\c ca de frequ\^encia depende da velocidade relativa fonte-observador, e tamb\'em da gravita\c c\~ao. Em particular, se ambos fonte luminosa e observador estiverem parados, e a gravita\c c\~ao for invariante ($g_{\mu\nu}$ n\~ao dependerem de $t$), n\'os calculamos o fator Doppler muito facilmente. De fato, neste caso o intertempo de coordenada $\dd t$ entre os sinais emitidos e os recebidos s\~ao o mesmo \cite[p\'ag.\,416]{Anderson}. Como para os dois korpos parados ocorre $\dd\tau_a=\sqrt{g_{00}(a)}\,\dd t$ e $\dd\tau_b=\sqrt{g_{00}(b)}\,\dd t$\,, a (\ref{Dtau}) e a (\ref{Schw}) fazem (veja a figura~\ref{FigDopEst})}

\bea                                                   \label{DopSchw}%04
D_{a\rightarrow b}=
\sqrt{\frac{1-\rho/a}{1-\rho/b}}\,.  
\eea

\ppln{Se $a<b$ do $D<1$, indikante ru^gi^gon; kaj se $a>b$ do $D>1$, indikante violi^gon. Se $a=b$ do $D=1$, indikante neeston de Dopplera efiko. Atentu, ke la angulaj pozicioj de kaj fonto kaj observanto {\it restantaj} ne gravas por~(\ref{DopSchw}).}
\pprn{Se $a<b$ ent\~ao $D<1$, indicando desvio para o vermelho; e se $a>b$ ent\~ao $D>1$, indicando desvio para o violeta. Se $a=b$ ent\~ao $D=1$, indicando inexist\^encia de efeito Doppler. Atente que as posi\c c\~oes angulares da fonte e do receptor {\it parados} n\~ao importam para a (\ref{DopSchw}).} 

\ppsection[0.6ex]{Geodezaj ekvacioj}{Equa\c c\~oes geod\'eticas}

\ppln{^Car la metrikaj koeficientoj de (\ref{Schw}) ne pendas de $t$, tial la tempa geodeza ekvacio generas $g_{00}\dd t/\dd\tau= {\rm konst}$, kaj do}
\pprn{Como os coeficientes m\'etricos de (\ref{Schw}) n\~ao de\-pen\-dem de $t$, a equa\c c\~ao geod\'etica temporal gera $g_{00}\dd t/\dd\tau= {\rm konst}$, e portanto}  

\bea                                                        \label{dt}%05 
\dd\tau_{geod}=\frac{1}{E}(1-\rho/r)\,\dd t\,, 
\hspace{3mm} E={\rm konst}\,\mathbf{.} 
\eea 

\ppln{^Car ili anka^u ne pendas de $\varphi$, tial la azimuta geodeza ekvacio generas $g_{\varphi\varphi}\dd\varphi/\dd\tau={\rm konst}$, kaj do, en ebeno $\theta=\pi/2$\,,}
\pprn{Como eles tampouco dependem de $\varphi$, a equa\c c\~ao geod\'etica azi\-mutal gera $g_{\varphi\varphi}\dd\varphi/\dd\tau={\rm konst}$, e portanto, no plano $\theta=\pi/2$\,,}   

\bea                                                      \label{dphi}%06
\dd\varphi_{geod}=
\frac{c\rho h}{r^2}\,(1-\rho/r)\,\dd t\,, \hspace{3mm} h={\rm konst}\,. 
\eea %=\frac{c\rho Eh}{r^2}\,\dd\tau_{geod}

\ppln{Konstantoj $E$ kaj $h$ estas nedimensiaj, kaj kvalite priskribas iun ajn geodezan movadon de materio (movadon de tempa tipo) en ebeno $\theta=\pi/2$\,. Fakte, iuj ajn du tiel trajektorioj kun samaj valoroj de $E$ kaj $h$ estas surmeteblaj per spaca rotacio en la ebeno.}
\pprn{As constantes $E$ e $h$ s\~ao adimensionais, e des\-crevem qualitativamente qualquer movimento geod\'etico de mat\'eria (movimento tipo tempo) no plano $\theta=\pi/2$\,. De fato, quaisquer duas tais trajet\'orias com mesmos valores de $E$ e $h$ s\~ao superpon\'iveis mediante rota\c c\~ao espacial no plano.}

\ppl{Fine, por venonta uzo en sekcio 5 ni prezentas la duagrade diferencialan geo\-dezan ekvacion por $r$\,:}
\ppr{Finalmente, para uso futuro na se\c c\~ao 5 n\'os apresentamos a equa\c c\~ao geod\'etica diferencial de ordem 2 para $r$\,:}  
%ERA $\frac{\dd ^2r}{\dd\tau^2} =
%\frac{\rho(\dd r/\dd\tau)^2}{2r^2(1-\rho/r)} - 
%\frac{E^2}{2r^3(1-\rho/r)}\left(\rho c^2r-2h^2(1-\rho/r)^2\right)\,;$

\bea                                                   \label{eqgeodr}%07
\frac{\dd^2r}{\dd\tau^2} -
\frac{\rho}{2r^2(1-\rho/r)}\left(\frac{\dd r}{\dd\tau}\right)^2 = 
r(1-\rho/r)\left(\frac{\dd\varphi}{\dd\tau}\right)^2 - 
\frac{\rho c^2(1-\rho/r)}{2r^2}\left(\frac{\dd t}{\dd\tau}\right)^2\,.
\eea
%essa equa\c c\~ao ser\'a usada no estudo de geod\'esicas circulares, na se\c c\~ao~{\bf\ref{geodcirk}}. 

\ppl{Por kompreni la rolon de konstanto $E$ en (\ref{dt}), konsideru normhorlo^gon movi\^gantan {\it geodeze}. ^Gi pasas preter punkto en radiusa pozicio $r$ en loka momento $t$, kaj pasas preter najbara punkto en loka momento $t+\dd t$\,. Do ^gia montrilo anta^ueniras d$\tau_{geod}$ kiel en (\ref{dt}), inter la du pozicioj. Atentu, ke la rilato $\dd\tau_{geod}/\dd t$ ne pendas de la direkto de movado de horlo^go.}
\ppr{Para compreender o papel da constante $E$ na (\ref{dt}), considere um rel\'ogio padr\~ao movendo-se {\it geodeticamente}. Ele passa por um ponto na posi\c c\~ao radial $r$ no momento local $t$, e passa por um ponto vizinho no momento local $t+\dd t$\,. Ent\~ao seu marcador avan\c ca d$\tau_{geod}$ como na (\ref{dt}), entre as duas posi\c c\~oes. Atente que a rela\c c\~ao $\dd\tau_{geod}/\dd t$ n\~ao depende da dire\c c\~ao do movimento do rel\'ogio.}   

\ppl{Praktika esprimo de konstanto $E$ estas havita el (\ref{Schw}) kaj (\ref{dt}),}
\ppr{Uma express\~ao pr\'atica da constante $E$ \'e obtida de (\ref{Schw}) e (\ref{dt}),} 

\bea                                                    \label{Egeral}%08 
E=\sqrt{\frac{1-\rho/r}{1-v^2/c^2}}\,, 
\eea 

\ppln{kie $v$ estas funkcio pendanta de movado (geodeza a^u ne) de korpo en ebeno $\theta=\pi/2$\,,}
\pprn{sendo $v$ uma fun\c c\~ao que depende do movimento (geod\'etico ou n\~ao) do corpo no plano $\theta=\pi/2$\,,} 

\bea                                                       \label{vel}%09
v:=\sqrt{\frac{(\dd r/\dd t)^2}{(1-\rho/r)^2} + \frac{r^2(\dd\varphi/\dd t)^2} {1-\rho/r}}\,. 
\eea

\ppln{Ni ofte perceptos, ke tiu $v$ tre bone ^generaligas la skalaran rapidon de speciala relativeco, kaj la skalaran Newtonan rapidon de korpo, en ebeno $\theta=\pi/2$. Ekzemple, la Newtona rapido de {\it eskapo} ekde distanco $r$ de centra maso $m$ estas bonkonate $\sqrt{2Gm/r}$\,; kaj la responda relativeca, ekde la radiusa pozicio $r$, estas havebla el (\ref{Egeral}) kun $E=1$, kaj estas $v= c\sqrt{\rho/r}\equiv\sqrt{2Gm/r}$. Do, ni nomos $v$ kiel rapido. Vere, $v$ rezultas el pli ^generala formulo, kiun ni prezentos en estonta artikolo~\cite{reltemp2}.}
\pprn{N\'os frequentemente vamos perceber que esse $v$ generaliza muito bem a velocidade escalar da rela\-tividade especial, e a velocidade escalar Newtoniana de um corpo, no plano $\theta=\pi/2$. Por exemplo, a velocidade Newtoniana de {\it escape} a partir da dist\^ancia $r$ \`a massa central $m$ \'e bem sabido $\sqrt{2Gm/r}$\,; e a correspondente relativista, a partir da posi\c c\~ao radial $r$, \'e obten\'ivel de (\ref{Egeral}) com $E=1$, e d\'a $v= c\sqrt{\rho/r}\equiv\sqrt{2Gm/r}$.  Assim, n\'os vamos chamar $v$ de velocidade. Em verdade, $v$ decorre de uma f\'ormula mais geral, que vamos apresentar em um futuro artigo~\cite{reltemp2}.}

\ppl{Atentu en (\ref{Egeral}), ke se korpo kun geodeza movado atingas  $r\rightarrow\infty$ kun $v_\infty\neq0$, do  $E=1/\sqrt{1-{v_\infty}^2/c^2}$, kaj do $E>1$; anka^u atentu, ke restado en senlima radiusa pozicio $(v_\infty=0)$ estas geodezo kun $E=1$. Kaj se geodeze movi^ganta korpo momente restas en iu limhava radiusa pozicio $r=b$, do $E=\sqrt{1-\rho/b}$\,, kaj do $E<1$\,; tiu pozicio estas tio de maksimuma distanco de korpo al centra maso.}
\ppr{Atente na (\ref{Egeral}) que, se um corpo com movimento geod\'etico atinge $r\rightarrow\infty$ com $v_\infty\neq0$, ent\~ao $E=1/\sqrt{1-{v_\infty}^2/c^2}$, portanto $E>1$; atente tamb\'em que o estado de repouso no infinito radial $(v_\infty=0)$ \'e uma geod\'esica com $E=1$. E se o corpo em movimento geod\'etico p\'ara momentaneamente em alguma posi\c c\~ao radial finita $r=b$, ent\~ao $E=\sqrt{1-\rho/b}$\,, portanto $E<1$\,; essa posi\c c\~ao \'e a de m\'axima dist\^ancia do corpo \`a massa central.} 

\ppl{Interesas rimarki, ke esprimo $E-1$ similas al Newtona totala energio. Fakte, en Newtona gravito ni difinas la totalan energion $E_t$ de korpo kiel la algebra adicio de \^gia kinetika energio (pozitiva) kun \^gia potenciala energio (malpozitiva). Se $E_t\geq0$ ($E\geq1$ relativece), la korpo povas fori^gi ^gis  $\infty$, tamen se $E_t<0$ ($0<E<1$ relativece), la kor\-po trakuras nur regionoj kun $r<\infty$.}
\ppr{\'E interessante notar que a express\~ao $E-1$ se assemelha \`a energia total Newtoniana. De fato, na gravita\c c\~ao Newtoniana n\'os definimos a energia total $E_t$ de um corpo como a soma alg\'ebrica de sua energia cin\'etica (positiva) com sua energia potencial (negativa). Se $E_t\geq0$ ($E\geq1$, relativistamente), o corpo pode se afastar at\'e o $\infty$, enquanto que se $E_t<0$ ($0<E<1$, relativistamente), o corpo percorre apenas regi\~oes com $r<\infty$.}

\ppsection[0.6ex]{Cirklaj geodezoj}{Geod\'esicas circulares} \label{geodcirk}

\ppln{Por studi cirklajn geodezojn de {\Sch a} metriko (\ref{Schw}) en ebeno $\theta=\pi/2$, ni konsideras  $r={\rm konst}$ en (\ref{eqgeodr}) kaj ricevas la tempe konstantan rotacirapidon $\omega:=\dd\varphi/\dd t$\,, }
%(((TIREI $\dd ^2r/\dd t^2=0$ kaj $\dd r/\dd t=0$,))) 
\pprn{Para estudar as geod\'esicas circulares da m\'etrica de \Sch\ (\ref{Schw}) no plano $\theta=\pi/2$, n\'os consideramos a (\ref{eqgeodr}) com $r={\rm konst}$, e obtemos a velocidade angular temporalmente constante $\omega:=\dd\varphi/\dd t$\,,} 
%(((TIREI $\dd ^2r/\dd t^2=0$ e $\dd r/\dd t=0$)))

\bea                                                    \label{wgeodc}%10
\omega_{geod}^{cirk}(r)=\pm\, c\sqrt{\frac{\rho}{2\,{r}^3}}\,,
\hspace{5mm} r={\rm konst}\,;
\eea 

\ppln{^car $\rho=2Gm/c^2$ tial ni konstatas, ke $\omega_{geod}^{cirk}(r)$ formale koincidas kun la Newtona geodeza rezulto $\pm\sqrt{Gm/r^3}$. Vidu kurbon $G$ en figuro~\ref{FigAB}.}
\pprn{como $\rho=2Gm/c^2$\,, n\'os constatamos que $\omega_{geod}^{cirk}(r)$ coincide formalmente com o resultado geod\'etico Newtoniano $\pm\sqrt{Gm/r^3}$. Veja a curva $G$ na figura~\ref{FigAB}.} 

\ppl{Uzante (\ref{dphi}) kaj (\ref{wgeodc}) ni vidas, ke la konstanto $h$ de cirkla geodeza movado valoras}
\ppr{Usando (\ref{dphi}) e (\ref{wgeodc}) vemos que a constante $h$ do movimento circular geod\'etico vale} 

\bea                                                     \label{hgeod}%11
h_{geod}^{cirk}(r)=\pm\,\frac{\sqrt{r/(2\rho)}}{1-\rho/r}\,,
\hspace{5mm} r={\rm konst}, 
\eea

\ppln{kaj uzante (\ref{Schw}) kaj (\ref{dt}) kaj (\ref{hgeod}), ni ricevas}  
\pprn{e usando (\ref{Schw}) e (\ref{dt}) e (\ref{hgeod}) n\'os recebemos}

\bea                                                     \label{Egeod}%12
E_{geod}^{cirk}(r)=\frac{1-\rho/r}{\sqrt{1-3\rho/(2r)}}\,,
\hspace{5mm} r={\rm konst}. 
\eea 

\ppln{Atentu en (\ref{Egeod}), ke cirkla geodezo de korpo havas   $E_{geod}^{cirk}\!<\!1$, kaj ekzistas nur se $r\!>\!3\rho/2$.} 
\pprn{Atente na (\ref{Egeod}) que uma geod\'esica circular de um corpo tem $E_{geod}^{cirk}\!<\!1$\,, e existe somente se $r\!>\!3\rho/2$\,.} 

\ppl{Fine, uzante (\ref{vel}) kun $\dd r/\dd t=0$\,, kaj  (\ref{wgeodc}), ni ricevas la tangentan rapidon de geodeza cirkla movado kun radiuso $r$:} 
\ppr{Finalmente, usando a (\ref{vel}) com $\dd r/\dd t=0$\,, e a (\ref{wgeodc}), obtemos a velocidade tangencial do movimento circular geod\'etico com raio $r$:} 

\bea                                                   \label{vgeocir}%13
v_{geod}^{cirk}(r)=c\sqrt{\frac{\rho}{2(r-\rho)}}\,,
\hspace{5mm} r={\rm konst}. 
\eea 

\ppln{Ni vidas, ke $v_{geod}^{cirk}(r)\rightarrow c$ se $r\rightarrow 3\rho/2$. Tio signifas, ke nur lumo movi^gas geodeze en cirklo kun tiu radiuso.} 
\pprn{Vemos que $v_{geod}^{cirk}(r)\rightarrow c$ se $r\rightarrow 3\rho/2$. Isso significa que somente a luz se move geodeticamente no c\'irculo com esse raio.} 

\ppsection[0.6ex]{Cirklaj movadoj}{Movimentos circulares}

\begin{figure}
\centerline{\epsfig{file=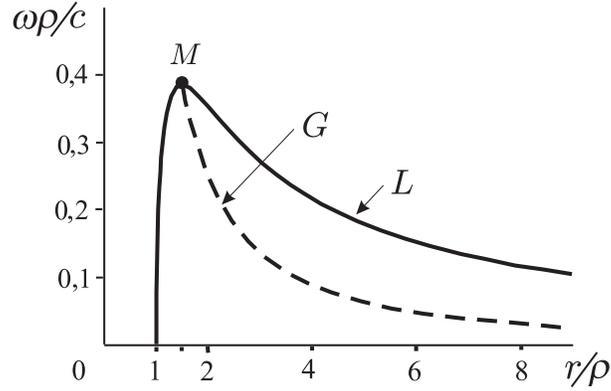,width=8cm}}
\selectlanguage{esperanto}\caption{\selectlanguage{esperanto}
Strekata kurbo $G$ montras la rotacirapidon $\omega$ de korpo kun cirkla geodeza movado kun radiuso $r$\,, kiel (\ref{wgeodc}). Plena kurbo $L$ montras la plej grandan rotacirapidon $\omega$ de korpo en cirkla movado kun radiuso $r$, kiel (\ref{wlumo}). Maksimumo punkto $M=[3/2, 2/\sqrt{27}]$ de kurbo $L$ indikas bonkonatan geodezan cirklan movadon de lumo.  
\ppdu{\newline \selectlanguage{portuguese}Figura~\ref{FigAB}:
A curva tracejada $G$ mostra a velocidade angular $\omega$ de um corpo com movimento circular geod\'etico com raio $r$\,, como a (\ref{wgeodc}). A curva cheia $L$ mostra a m\'axima velocidade angular $\omega$ de um corpo em movimento circular com raio $r$\,, como a (\ref{wlumo}). O ponto m\'aximo $M=[3/2, 2/\sqrt{27}]$ da curva $L$ indica o bem conhecido movimento geod\'etico circular de luz.}}
\label{FigAB}
\end{figure}

\ppln{Ni studos movadojn, geodezajn a^u ne, kun $\theta=\pi/2$ kaj $r={\rm konst}$, en {\Sch a} metriko (\ref{Schw}). Malsimile al la Newtona kinematiko, la relativeca teorio limigas la rotacirapidon de tiuj movadoj. Fakte, en cirkla movado kun radiuso $r$, la kondi^co $\dd s^2>0$ por movado de korpo faras, ke la rotacirapido $\omega:=\dd\varphi/\dd t$ havu modulon plieta ol}
\pprn{Vamos estudar movimentos, geod\'eticos ou n\~ao, com $\theta=\pi/2$ e $r={\rm konst}$, na m\'etrica de \Sch\ (\ref{Schw}). Diferentemente da cinem\'atica Newtoniana, a relatividade limita a velocidade angular desses movimentos. De fato, em um movimento circular com raio $r$, a condi\c c\~ao $\dd s^2>0$ para movimento de um corpo faz com que a velocidade angular $\omega:=\dd\varphi/\dd t$ tenha m\'odulo menor que} 

\bea                                                     \label{wlumo}%14
\omega_{limo}(r):=\frac{c}{r}\,\sqrt{1-\rho/r}\,. 
\eea 

\ppln{Tio estus la $\omega$ de lumo voja^ganta kun rapido $c$ en cirklo kun radiuso $r$. Kurbo $L$ en figuro~\ref{FigAB} montras la interesan funkcion (\ref{wlumo}). Atentu, ke paroj $[r, \omega]$ donante cirklan movadon de korpo ekzistas nur sub kurbo $L$. Tiu kurbo anka^u montras, ke duo de valoroj de radiuso $r$ de la cirklo havas saman $\omega_{limo}$. Por la sola radiuso $r=3\rho/2$\,, $\omega_{limo}(r)$ atingas maksimuman valoron }
\pprn{Esse seria o $\omega$ de luz viajando com velocidade $c$ em um c\'irculo com raio $r$. A curva $L$ na figura~\ref{FigAB} mostra a interessante fun\c c\~ao (\ref{wlumo}). Atente que pares $[r, \omega]$ dando movimentos circulares de um corpo existem somente abaixo da curva $L$\,. Essa curva mostra tamb\'em que um par de valores do raio $r$ do c\'irculo t\^em mesmo $\omega_{limo}$. Para o \'unico raio $r=3\rho/2$\,, $\omega_{limo}(r)$ atinge um m\'aximo valor } 

\bea                                                      \label{wmax}%15
\omega_{maks}:=\frac{2}{\sqrt{27}}\,\frac{c}{\rho}\,. 
\eea 

\ppln{Vidu punkton $M=[3/2, 2/\sqrt{27}]$ en figuro~\ref{FigAB}\,. ^Car (\ref{wgeodc}) kun $r=3\rho/2$ faras $|\omega_{geod}^{cirk}|=(2/\sqrt{27})(c/\rho)$ same kiel (\ref{wmax}), tial punkto $M$ indikas {\it luman geodezon.} 
Indas komenti, ke en ordinaraj fizikaj sistemoj tiu relativeca limigo   $|\omega|\leq\omega_{maks}$ ordinare ne gravas, ^car la materio generanta graviton ordinare okupas tiun radiusan pozicion $3\rho/2$.}  
\pprn{Veja o ponto $M=[3/2, 2/\sqrt{27}]$ na figura~\ref{FigAB}\,. Como (\ref{wgeodc}) com $r=3\rho/2$ faz $|\omega_{geod}^{cirk}|=(2/\sqrt{27})$ $(c/\rho)$ como a (\ref{wmax}), o ponto $M$ indica {\it geod\'esica de luz}. Vale a pena comentar que em sistemas f\'isicos usuais essa limita\c c\~ao re\-la\-tivista  $|\omega|\leq\omega_{maks}$ geralmente n\~ao importa, porque a mat\'eria gera\-dora da gravita\c c\~ao usualmente ocupa essa  posi\c c\~ao radial $3\rho/2$.}

\begin{figure}
\centerline{\epsfig{file=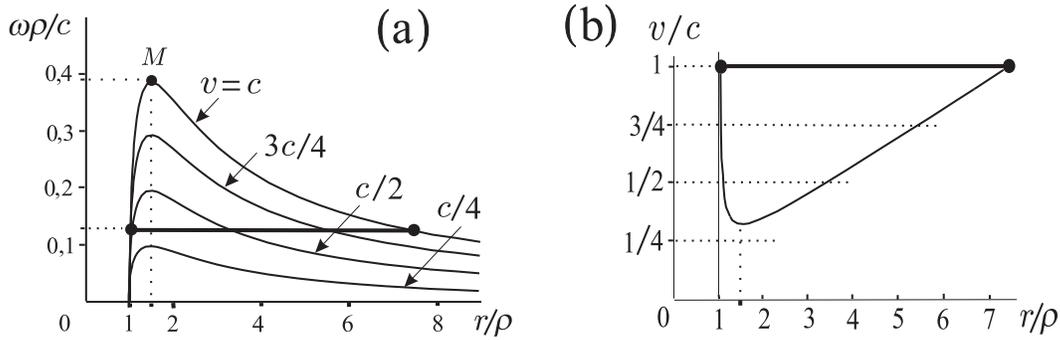,width=14cm}}
\selectlanguage{esperanto}\caption{\selectlanguage{esperanto}
(a) ^Ciu kurbo montras parojn $[r, \omega]$ por cirkla movado kun la indikata tangenta rapido. La horizonta streko indikas stangon rotaciantan kun rotacirapido $\omega=0,125\,c/\rho$ ^cirka^ue la centra maso, kaj ku^sante inter radiusaj pozicioj $r_{min}=1,016\rho$ kaj $r_{max}=7,25\rho$\,. (b) Rapido (\ref{velcirk}) de punktoj de stango en anta^ua figuro. Amba^u ekstremoj havas rapidon $v\rightarrow c$, kaj la pli eta rapido okazas en $r=3\rho/2$\,.    
\ppdu{\newline \selectlanguage{portuguese}Figura~\ref{FigRapidoj}:
(a) Cada curva mostra pares $[r, \omega]$ para movimento circular com a velocidade tangencial indicada. O tra\c co horizontal indica uma barra rodando com $\omega=0,125\,c/\rho$ em torno da massa central, e deitada entre as posi\c c\~oes radiais $r_{min}=1,016\rho$ e $r_{max}=7,25\rho$\,. (b) Velocidade (\ref{velcirk}) dos pontos da barra na figura anterior. Ambas extremidades t\^em velocidade $v\rightarrow c$, e a menor velocidade ocorre em $r=3\rho/2$\,.}}
\label{FigRapidoj}
\end{figure}

\ppl{Newtone, stango povas rotacii kun iu ajn rotacirapido $\omega$ ^cirka^u unu el ^giaj ekstremoj, kaj povas esti iom ajn longa. Tamen special relativeco limigas la longon al maksimuma valoro $c/\omega$ tial, ke la ekstera ekstremo ne havu rapidon pli granda ol $c$. Plua limigo estas ^ce \Sch\ tial, ke la ena ekstremo ne havu rapidon $c$\,.} 
\ppr{Newtonianamente uma barra pode rodar com qualquer $\omega$ em torno de uma das extremidades, e pode ter qualquer comprimento. Por\'em a rela\-tividade especial limita o comprimento ao valor m\'aximo $c/\omega$, para que a extremidade exterior n\~ao tenha velocidade maior que $c$. Uma limita\c c\~ao adicional existe em \Sch, para que a extremidade interior n\~ao tenha velocidade $c$\,.} 

\ppl{Fakte, la tangenta rapido de punkto $r$ de stango kun rotacirapido $\omega$ estas} 
\ppr{De fato, a velocidade tangencial de um ponto $r$ da barra com velocidade angular $\omega$ \'e} 

\bea                                                   \label{velcirk}%16
v_{cirk}(r)=\frac{r\omega}{\sqrt{1-\rho/r}}\,, 
\eea 

\ppln{do la limigo $v<c$ (ekvivalente $\dd s^2>0$ en (\ref{Schw})) generas limigojn $r_{min}(\omega)<r<r_{max}(\omega)$\,. Figuro~\ref{FigRapidoj}.a montras kurbojn kies paroj $[r, \omega]$ havas la indikatan tangentan rapidon $v$. ^Gi anka^u montras stangon radiuse ku^santan inter radiusaj pozicioj $r_{min}$ kaj $r_{max}$, kiun rotacias kun $\omega=0,125\,c/\rho$\,. Figuro~\ref{FigRapidoj}.b montras rapidon (\ref{velcirk}) de punktoj de stango. } 
\pprn{ent\~ao a limita\c c\~ao $v<c$ (equivalentemente $\dd s^2>0$ na (\ref{Schw})) gera as limita\c c\~oes $r_{min}(\omega)<r<r_{max}(\omega)$\,. A figura~\ref{FigRapidoj}.a mostra curvas cujos pares $[r, \omega]$ t\^em a velocidade tangencial $v$ indicada. Ela mostra tamb\'em uma barra estendida radialmente entre as posi\c c\~oes radiais $r_{min}$ e $r_{max}$\,, e rodando com velocidade angular $\omega=0,125\,c/\rho$\,. A figura~\ref{FigRapidoj}.b mostra a velocidade (\ref{velcirk}) dos pontos da barra. } 

\ppsection[0.6ex]{Komparo de propratempoj -- cirkle}{Compara\c{c}\~ao de tempos pr\'oprios -- circular}
\label{CIPI}

\ppln{Ni komparas la tempan fluon de du normhor\-lo^goj malsimile movi^gantaj. La unua movi\-^gas cirkle kun radiuso $r$ kaj rotacirapido $\dd\varphi/\dd t=\omega={\rm konst}$, supozata pozitiva. La^u (\ref{Schw}), la propra intertempo post unu kompleta turno estas}
\pprn{Vamos comparar o fluir do tempo de dois rel\'ogios padr\~ao em diferentes movimentos. O primeiro tem movimento circular com raio $r$ e velocidade angular $\dd\varphi/\dd t=\omega={\rm konst}$, suposta  positiva. Segundo a (\ref{Schw}), o intertempo pr\'oprio ap\'os uma volta completa \'e} 

\bea                                                   \label{Dtaumov}%17
\Delta\tau_{mov}= (2\pi/\omega)\sqrt{1-\rho/r-\omega^2r^2/c^2}\,. 
\eea 

\ppln{La alia normhorlo^go, restanta en tiu radiusa pozicio $r$ dum la sama koordinata intertempo $\Delta t=2\pi/\omega$, montras propran intertempon}
\pprn{O outro rel\'ogio padr\~ao, parado naquela posi\c c\~ao radial $r$ durante o mesmo intertempo de coordenada $\Delta t=2\pi/\omega$, mostra um intertempo pr\'oprio} 

\bea                                                   \label{Dtaurip}%18
\Delta\tau_{rip}=(2\pi/\omega)\sqrt{1-\rho/r}\,. 
\eea 

\ppln{Esprimoj (\ref{Dtaumov}) kaj (\ref{Dtaurip}) simpli^gas se ni uzas $v=\omega r/\sqrt{1-\rho/r}$, la rapidon (\ref{velcirk}) de cirkle movi^ganta horlo^go. Tiuokaze,}
\pprn{As express\~oes (\ref{Dtaumov}) e (\ref{Dtaurip}) se simplificam se usar\-mos $v=\omega r/\sqrt{1-\rho/r}$, a velocidade (\ref{velcirk}) do rel\'ogio em movimento circular. Nesse caso,}

\bea                                                        \label{DD}%19
\Delta\tau_{mov}= \frac{2\pi r}{v\gamma}\,, \hspace{3mm} \Delta\tau_{rip}= \frac{2\pi r}{v}\,, 
\eea 

\ppln{estante $\gamma:=1/\sqrt{1-v^2/c^2}$\,. ^Car $\gamma>1$,  tial $\Delta\tau_{mov}<\Delta\tau_{rip}$\,, tio estas, la tempo de la movi^ganta normhorlo^go fluis pli malrapide ol la tempo de la restanta normhorlo^go.}
\pprn{sendo $\gamma:=1/\sqrt{1-v^2/c^2}$\,. Como $\gamma>1$, ent\~ao $\Delta\tau_{mov}<\Delta\tau_{rip}$\,, isto \'e, o tempo do rel\'ogio em movimento fluiu mais lentamente do que o tempo do rel\'ogio parado.} 

\ppl{Tiu rezulto koincidas kun tio de special-relativeco, ke  $\Delta\tau$ de movi^ganta horlo^go estas plieta ol tio de restanta horlo^go, poste unu kompleta turno. Certe, la fakto ke la du horlo^goj estas en gravita potencialo kun sama valoro estas baza por tiu koincido. Krome, ne gravas se la movado estas geodeza, a^u ne. }  
\ppr{Esse resultado coincide com o da relatividade especial, que o $\Delta\tau$ de um rel\'ogio em movimento \'e menor que o de um rel\'ogio parado, ap\'os uma volta completa. Certamente o fato de os dois rel\'ogios estarem em potencial gravitacional com mesmo valor \'e fundamental para essa coincid\^encia. Al\'em disso, n\~ao \'e importante se o movimento \'e geod\'etico, ou n\~ao. }

\ppl{El ekvacio (\ref{Dtaumov}) oni povas kalkuli Doppleran faktoron inter du korpoj, cirkle movi^gantaj en la sama ebeno, kun sama konstanta rotacirapido kaj malsamaj radiusoj. Fakte, dividante (\ref{Dtaumov}) per ^gi mem kun malsamaj radiusoj, oni ricevas} 
\ppr{A partir da (\ref{Dtaumov}) pode-se calcular o fator Doppler entre dois corpos, movendo-se em c\'irculos no mesmo plano, com mesma velocidade angular constante e com raios diferentes. De fato, dividindo a (\ref{Dtaumov}) por ela mesma com raios diferentes se recebe}

\bea                                                     \label{Drarb}%20
D_{a\rightarrow b} =\sqrt{\frac{1-\rho/r_a-\omega^2{r_a}^2/c^2}{1-\rho/r_b-\omega^2{r_b}^2/c^2}}\,. 
\eea

\ppsection[0.6ex]{Kunaj cirklaj \ppdu{\\ }movadoj}{Movimentos circulares \\ conjugados}

\ppln{Konsideru du normhorlo^gojn cirkle mov\-i^gantajn kun sama rotacirapido $\omega:=\dd\varphi/\dd t={\rm konst}$ en ebeno $\theta=\pi/2$ de metriko (\ref{Schw}), sed en malsamaj radiusoj $r_1$ kaj $r_2$. La^u la Newtona kinematiko, la horlo^go en cirklo kun plieta radiuso havas plietan tangentan rapidon. La^u la special-relativeco, tiu fakto farus ke ^gia propratempo fluu {\it pli} rapide ol tiu de la alia. Sed ^gi trakuras regionojn kun plieta $g_{00}$; la^u la ^general-relativeco, tiu alia fakto farus ke ^gia propratempo fluu {\it malpli} rapide ol tiu de la alia. Ni demandas: ^cu la du efikoj povas ekzakte ekvilibri?} 
\pprn{Considere dois rel\'ogios padr\~ao em movimentos circulares com mesma velocidade angular $\omega:=\dd\varphi/\dd t={\rm konst}$ no plano $\theta=\pi/2$ da m\'etrica (\ref{Schw}), por\'em em raios $r_1$ e $r_2$ diferentes. Segundo a cinem\'atica Newtoniana o rel\'ogio no c\'irculo com raio menor tem menor velocidade tangencial. Segundo a relatividade especial, esse fato faria com que seu tempo pr\'oprio flu\'isse {\it mais} rapidamente que o do outro.  Por\'em ele percorre regi\~oes com menor $g_{00}$; segundo a relatividade geral, esse outro fato faria com seu tempo pr\'oprio flu\'isse {\it menos} rapidamente que o do outro. N\'os perguntamos: podem os dois efeitos exatamente compensar-se?} 
%\\ {\bf Heuristiko = scienco pri la metodoj de esplorado en la sciencoj.} 

\ppl{La respondo estas {\it jes}. Por ^ciu valoro de $\omega$, ni trovos ne-nombreblan kvanton da paroj de radiusoj  $(r_1,r_2)$ tiel ke la propraj intertempoj $\Delta\tau_1$ kaj $\Delta\tau_2$ de la horlo^goj estas samaj, post unu kompleta turno. La graveco de tiu rezulto estas, ke lumaj signaloj igitaj el unu horlo^go al la alia ne havas Doppleran efikon.} 
\ppr{A resposta \'e {\it sim}. Para cada valor de $\omega$, n\'os vamos encontrar uma quantidade n\~ao-numer\'avel de pares de raios $(r_1,r_2)$ tais que os intertempos pr\'oprios $\Delta\tau_1$ e $\Delta\tau_2$ dos dois rel\'ogios sejam iguais, ap\'os uma volta completa. A relev\^ancia desse resultado \'e que sinais luminosos emitidos de um rel\'ogio para o outro n\~ao t\^em efeito Doppler.}

\ppl{Fakte, ekvacio (\ref{Schw}) kun $\dd r=0$ faras $(\dd\tau_1)^2=(1-\rho/r_1)(\dd t)^2-(r_1^2/c^2)(\dd\varphi)^2$ kaj  $(\dd\tau_2)^2=(1-\rho/r_2)(\dd t)^2-(r_2^2/c^2)(\dd\varphi)^2$\,. Uzante $\dd\tau_1=\dd\tau_2$ kaj $\omega:=\dd\varphi/\dd t$ okazas} 
\ppr{De fato, a equa\c c\~ao (\ref{Schw}) com $\dd r=0$ produz $(\dd\tau_1)^2=(1-\rho/r_1)(\dd t)^2-(r_1^2/c^2)(\dd\varphi)^2$ e igualmente $(\dd\tau_2)^2=(1-\rho/r_2)(\dd t)^2-(r_2^2/c^2)(\dd\varphi)^2$\,. Usando $\dd\tau_1=\dd\tau_2$ e $\omega:=\dd\varphi/\dd t$ ocorre}
% h\'a ainda a solu\c c\~ao $r_1=r_2$  

\bea                                                       \label{r2}%21
r_1r_2(r_1+r_2)=\frac{\rho c^2}{\omega^2}\,.
\eea

\ppln{Atentu, ke oni anka^u povus ricevi (\ref{r2}) farante $D_{a\rightarrow b}=1$ en (\ref{Drarb}). Ekvacio (\ref{r2}) estas simetria per ^san^go $r_1\leftrightarrow r_2$\,. Solvata, ^gi prezentas radiuson $r_2$ kiel funkcio de radiuso $r_1$ la^u}
\pprn{Atente que pode-se ainda receber a (\ref{r2}) fazendo $D_{a\rightarrow b}=1$ na (\ref{Drarb}). A equa\c c\~ao (\ref{r2}) \'e sim\'etrica perante a troca $r_1\leftrightarrow r_2$\,. Resolvida, ela apresenta o raio $r_2$ como fun\c c\~ao do raio $r_1$ segundo}

\bea                                                       \label{r3}%22
r_2=\frac{r_1}{2}\left(\sqrt{1+\frac{4c^2\rho}{\omega^2r_1^3}}-1\right)\,.
\eea

\ppl{Uzante (\ref{wgeodc}) en (\ref{r2}) ni konstatas, ke se movado kun radiuso $r_1$ estas geodeza, do $r_2=r_1$. Ni anka^u atentas en (\ref{r2}), ke se $r_1$ plii^gas, do $r_2$ plieti^gas, kaj reciproke. Oni konkludas ke, se $r_1$ estas plieta ol la geodeza radiuso, do $r_2$ estas pligranda ol la geodeza radiuso. Figuro~\ref{FigPares} remontras kurbojn $L$ kaj $G$ el figuro~\ref{FigAB} kaj iom da paroj de radiusoj por speciala rotacirapido $\omega=0,125\,c/\rho$\,. Oni povas montri, ke rapido (\ref{velcirk}) de korpo en ekstera cirklo estas ^ciam pligranda ol de korpo en ena cirklo, same kiel la parolado de la unua paragrafo de ^ci tiu sekcio.}  
\ppr{Usando a (\ref{wgeodc}) na (\ref{r2}) n\'os constatamos que se o movimento com raio $r_1$ for geod\'etico, ent\~ao $r_2=r_1$. N\'os tamb\'em vemos na (\ref{r2}) que se $r_1$ cresce, ent\~ao $r_2$ diminui, e reciprocamente. Conclui-se que, se $r_1$ for menor que o raio geod\'etico, ent\~ao $r_2$ ser\'a maior que o raio geod\'etico. A figura~\ref{FigPares} novamente mostra as curvas $L$ e $G$ da figura~\ref{FigAB} e alguns dos pares de raios para a particular velocidade angular $\omega=0,125\,c/\rho$\,. Pode-se mostrar que a velocidade (\ref{velcirk}) do corpo no c\'irculo externo \'e sempre maior que a do corpo no c\'irculo interno, como na discuss\~ao do primeiro par\'agrafo desta se\c c\~ao.}
%Maple 091104a

\begin{figure}
\centerline{\epsfig{file=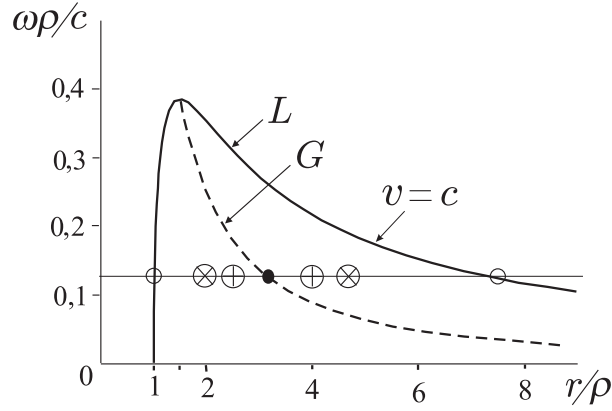,width=8cm}}
\selectlanguage{esperanto}\caption{\selectlanguage{esperanto}
Kurboj $L$ kaj $G$ el figuro~\ref{FigAB} kaj kvar paroj ($\circ, \otimes, \oplus, \bullet$) de korpoj en radiusoj $(r_1\,, r_2)$ el ekvacio (\ref{r3}), kun rotacirapido $\omega=0,125\,c/\rho$; por ^ciu paro, la propratempoj de korpoj fluas same. Speciale, atentu liman paron ($\circ$) kun rapido $c$\,, kaj atentu paron $r_1=r_2$ ($\bullet$) kun geodeza movado. 
\ppdu{\newline\selectlanguage{portuguese}Figura~\ref{FigPares}:
Curvas $L$ e $G$ da figura~\ref{FigAB} e quatro pares ($\circ, \otimes, \oplus, \bullet$) de corpos nos raios $(r_1\,, r_2)$ da equa\c c\~ao (\ref{r3}), com velocidade angular $\omega=0,125\,c/\rho$; para cada par, os tempos pr\'oprios dos corpos fluem igualmente. Em particular, atente para o par limite ($\circ$) com velocidade $c$\,, e atente para o par  $r_1=r_2$ ($\bullet$) com movimento geod\'etico.}}
\label{FigPares}
\end{figure}

\ppsection[0.6ex]{Komparo de propratempoj -- radiuse}{Compara\c{c}\~ao de tempos pr\'oprios -- radial}

\ppln{Normhorlo^go komence ripozanta en $r=a>\rho$ estas subite ^jetata radiuse, eksteren, en geodeza movado. ^Gi movi^gas ^gis la maksimuma  radiusa pozicio $r=b>a$, kaj revenas al pozicio $r=a$. Alia normhorlo^go restas en radiuso $r=a$ dum la tuta tempo.}
\pprn{Um rel\'ogio padr\~ao inicialmente em repouso em $r=a>\rho$ \'e subitamente jogado para fora radialmente, em movimento geod\'etico. Ele se move at\'e a m\'axima posi\c c\~ao radial $r=b>a$, e retorna \`a posi\c c\~ao $r=a$. Outro rel\'ogio padr\~ao fica parado no raio $r=a$ durante o tempo todo.} 

\ppl{Ni demandas: en renkonto, kiu horlo^go prezentas pligrandan varion  $\Delta\tau$ de propratempo? La fakto, ke la geodeza horlo^go movi^gas, dume la ne geodeza ^ciam restas, favoras ke $\Delta\tau_{geod}$ estu {\it plieta}. Sed la fakto, ke la geodeza horlo^go trakuras regionojn kun pligranda $g_{00}$ favoras ke $\Delta\tau_{geod}$ estu {\it pligranda}. Do la demando ne posedas tujan evidentan respondon.}
\ppr{N\'os perguntamos: no reencontro, qual rel\'ogio apresenta a maior varia\c c\~ao $\Delta\tau$ de tempo pr\'oprio? O fato que o rel\'ogio geod\'etico est\'a em movimento enquanto o n\~ao-geod\'etico est\'a sempre parado favorece $\Delta\tau_{geod}$ ser {\it menor}. Mas o fato que o rel\'ogio geod\'etico percorre regi\~oes com $g_{00}$ maior favorece $\Delta\tau_{geod}$ ser {\it maior}. Ent\~ao a pergunta n\~ao possui uma resposta imediata \'obvia.}  

\ppl{^Car la geodeze movi^ganta horlo^go momente restas en $r=b$, tial $v_b=0$\,, kaj do (\ref{Egeral}) faras $E=\sqrt{1-\rho/b}$ por ^gia movado. Tio en (\ref{dt}) okazigas $(1-\rho/b)(\dd\tau)^2 = (1-\rho/r)^2(\dd t)^2$\,, ke uzata en metriko (\ref{Schw}) faras (memoru, ke $\dd s^2=c^2(\dd\tau)^2$)}
\ppr{Como o rel\'ogio em movimento geod\'etico est\'a momentaneamente parado em $r=b$, ent\~ao $v_b=0$\,, e portanto a  (\ref{Egeral}) faz $E=\sqrt{1-\rho/b}$ para o seu movimento. Isso na (\ref{dt}) ocasiona $(1-\rho/b)(\dd\tau)^2 = (1-\rho/r)^2(\dd t)^2$\,, que usado na m\'etrica (\ref{Schw}) faz (relembre que $\dd s^2=c^2(\dd\tau)^2$)}

\bea                                                  \label{dtaugeod}%23
\frac{(\dd r)^2}{\rho/r-\rho/b}=(c\,\dd\tau_{geod})^2\,; 
\eea

\ppln{integrante duapotencan radikon de (\ref{dtaugeod}) de $r=a$ ^gis $r=b$ aperigas $\Delta\tau_{geod}(\rho,a,b)$:}
\pprn{integrando a raiz quadrada da (\ref{dtaugeod}) de $r=a$ at\'e $r=b$ revela $\Delta\tau_{geod}(\rho,a,b)$:} 

\bea                                                  \label{Dtaugeod}%24
c\Delta\tau_{geod}=\sqrt{b/\rho}\left(\sqrt{a(b-a)}+b\,{\rm cos}^{-1}\sqrt{a/b}\,\right)\,.
\eea

\ppl{Dum tiu intertempo, la montro de la normhorlo^go restanta en $r=a$ anta^ueniras, la^u sekcio~2, }
\ppr{Nesse intertempo, a marca\c c\~ao do rel\'ogio padr\~ao parado em $r=a$ avan\c ca, segundo a se\c c\~ao~2,} 

\bea                                                    \label{taurip}%25
\Delta\tau_{rip}=\sqrt{1-\rho/a}\,\Delta t\,, 
\eea

\ppln{estante $\Delta t$ la koordinata intertempo dum la forirado de la voja^ganta horlo^go. Simile al (\ref{dtaugeod}), pravi^gas}
\pprn{sendo $\Delta t$ o intertempo de coordenada durante o afastamento do rel\'ogio viajante. Semelhantemente \`a (\ref{dtaugeod}), ocorre}

\bea                                                       \label{oi1}%26
\frac{(\dd r)^2}{(1-\rho/r)^2(\rho/r-\rho/b)}=\frac{(c\dd t)^2}{1-\rho/b}\,, 
\eea

\ppln{kio interrilatas tempan kaj radiusan koordinatojn de movi^ganta horlo^go. Integrante la duapotencan radikon de (\ref{oi1}) de $r=a$ ^gis $r=b$ oni malkovras la intertempon $\Delta t$\,. Uzante tiun $\Delta t$ en (\ref{taurip}) rezultigas la propran intertempon de la ripozanta horlo^go} 
\pprn{que relaciona as coordenadas temporal e radial do rel\'ogio em movimento. Integrando a raiz quadrada da (\ref{oi1}) desde $r=a$ at\'e $r=b$ descobre-se o intertempo $\Delta t$\,. Usando este $\Delta t$ na (\ref{taurip}) faz resultar o intertempo pr\'oprio do rel\'ogio parado}

\bea                                                  \label{Dtaurip2}%27 
c\Delta\tau_{rip} = \sqrt{1\!-\!\rho/a}\left(\sqrt{b/\rho\!-\!1}\left(\sqrt{a(b\!-\!a)} + (b\!+\!2\rho)\,{\rm cos}^{-1}\sqrt{a/b}\,\right) + 2\rho\,{\rm tanh}^{-1}\!\sqrt{\frac{b/a-1}{b/\rho-1}}\,\right)\!.
\eea

\ppln{Atentu, ke se $b=a>\rho$ do ne estas movado, kaj do $\Delta\tau_{rip}=0$. Atentu anka^u, ke se la ejo de ^jeto estas tre proksima al la radiuso de \Sch, t.e., se $a\rightarrow\rho$\,, do $\Delta\tau_{rip}\rightarrow0$ kvankam $\Delta t\rightarrow\infty$\,.} 
\pprn{Atente que se $b=a>\rho$ ent\~ao n\~ao h\'a movimento, e portanto  $\Delta\tau_{rip}=0$. Atente tamb\'em que se o ponto de lan\c camento for muito pr\'oximo do raio de \Sch, i.e., se $a\rightarrow\rho$\,, ent\~ao $\Delta\tau_{rip}\rightarrow0$ embora $\Delta t\rightarrow\infty$\,.} 

\ppl{Intertempoj (\ref{Dtaurip2}) kaj (\ref{Dtaugeod}) estas nelonge prezentitaj per Gr\o n kaj Braeck~\cite{GronBraeck}. Nek ili nek ni sukcesis montri {\it algebre}, ke   $\Delta\tau_{geod}>\Delta\tau_{rip}$ por iuj ajn valoroj de $\rho<a<b$. Tamen, numeraj kalkuloj kaj tridimensiaj grafika^joj indikas tiun rezulton. Tio montras, ke la efikoj akumulitaj per gravito en ^ci tiu movado estas pli fortaj ol la efikoj akumulitaj per rapido, kontra^ue al okazo de sekcio~7.} %\ref{CIPI}.} 
% Maple (090527a.mws)
\ppr{Os intertempos (\ref{Dtaurip2}) e (\ref{Dtaugeod}) foram recentemente apresentados por Gr\o n e Braeck~\cite{GronBraeck}. Nem eles nem n\'os conseguimos mostrar {\it algebricamente} que  $\Delta\tau_{geod}>\Delta\tau_{rip}$ para quaisquer valores de $\rho<a<b$. Entretanto, c\'alculos num\'ericos e gr\'aficos tridimensionais indicam esse resultado. Isso mostra que os efeitos acumulados por gravita\c c\~ao neste movimento s\~ao mais fortes que os efeitos acumulados por velocidade, ao contr\'ario do caso da se\c c\~ao~7.}

\ppsection[0.6ex]{Dopplera faktoro \ppdu{\\ }en radiusaj movadoj}{Fator Doppler \\ em movimentos radiais}

\ppln{Luma fonto$\!$ movi^gas$\!$ radiuse$\!$ en spacotempo (\ref{Schw}), kaj eligas lumon el radiusa pozicio \mbox{$a>\rho$}. Tiu lumo estas perceptata per observanto restanta en radiusa pozicio $b\!>\!\rho$. Amba^u okazoj $a\!>\!b$ kaj $a\!<\!b$ estas pripensindaj. La angulaj valoroj $\theta$ kaj $\varphi$ de pozicioj de eligo kaj percepto estas samaj. Ni scivolas la Doppleran faktoron de percepto.}
\pprn{Uma fonte luminosa se move radialmente no espa\c cotempo (\ref{Schw}), e emite luz da posi\c c\~ao radial $a>\rho$. Essa luz \'e percebida por um observador parado na posi\c c\~ao radial $b>\rho$. Ambos casos  $a>b$ e $a<b$ devem ser considerados. Os valores angulares $\theta$ e $\varphi$ das posi\c c\~oes de emiss\~ao e recep\c c\~ao s\~ao iguais. Queremos saber o fator Doppler de recep\c c\~ao.}

\ppl{Same kiel en sekcio~3 ni faras, ke fonto eligu du sinsekvajn lumajn signalojn, en koordinatmomentoj $t_1$ kaj $t_1+\dd_at$\,, estante fonto en radiusaj pozicioj $a$ kaj $a+\dd_ar$, respektive; kaj $\dd_at>0$\,. ^Car la fonto movi^gas a^u eksteren a^u enen,  tial amba^u okazoj $\dd_ar>0$ kaj $\dd_ar<0$ estas konsiderindaj. Observanto {\it restas} en radiusa pozicio $b$, kaj ricevas la signalojn en koordinatmomentoj $t_2$ kaj $t_2+\dd_bt$, respektive. Por uzo en esprimo (\ref{Dtau}) de Dopplera faktoro, ni kalkulas la proprajn intertempojn $\dd\tau_a$ kaj $\dd\tau_b$\,, respondajn al $\dd t_a$ kaj $\dd t_b$\,, respektive.}
\ppr{Como na se\c c\~ao 3, n\'os fazemos com que a fonte emita dois sinais luminosos consecutivos, nos momentos de coordenada $t_1$ e $t_1+\dd_at$\,, estando a fonte nas posi\c c\~oes radiais $a$ e $a+\dd_ar$, respectivamente; e $\dd_at>0$\,. Como a fonte se move ou para fora ou para dentro, ambos casos $\dd_ar>0$ e $\dd_ar<0$ devem ser considerados. Um observador est\'a {\it parado} na posi\c c\~ao radial $b$, e recebe os sinais nos momentos de coordenada $t_2$ e $t_2+\dd_bt$, respectivamente. Para uso na express\~ao (\ref{Dtau}) do fator Doppler, n\'os vamos calcular os intertempos pr\'oprios $\dd\tau_a$ e $\dd\tau_b$\,, correspondentes a $\dd t_a$ e $\dd t_b$\,, respectivamente.}

\ppl{Uzante (\ref{Schw}) kaj (\ref{vel}) kun $\dd\varphi=0$ kaj $\dd\theta=0$ oni ricevas, por movi^ganta fonto,}
\ppr{Usando a (\ref{Schw}) e a (\ref{vel}) com $\dd\varphi=0$ e $\dd\theta=0$ se recebe, para a fonte em movimento,} 

\bea                                                     \label{dtaua}%28 
\dd\tau_a=\sqrt{1-{v_a}^2/c^2}\sqrt{1-\rho/a}\,\dd_at\,. 
\eea

\ppln{Simile, por la restanta observanto en radiusa pozicio $b$, la  intertempo $\dd\tau_b$ estas}
\pprn{Igualmente, para o observador parado na posi\c c\~ao radial $b$, o intertempo $\dd\tau_b$ \'e} 
\bea                                                     \label{dtaub}%29
\dd\tau_b=\sqrt{1-\rho/b}\,\dd_bt\,.%(\Delta_2t-\Delta_1t)\,. 
\eea 

\ppl{Ni bezonas rilatigi $\dd_bt$ al $\dd_at$\,. La ekvacio de movado de {\it luma signalo} en radiusa movado estas havebla farante  $\dd s^2=0$ en linia elemento (\ref{Schw}) kaj konsiderante $\dd\theta=0$ kaj $\dd\varphi=0$. Rezultas}
\ppr{Precisamos relacionar $\dd_bt$ a $\dd_at$\,. A equa\c c\~ao do movimento de um {\it sinal luminoso} em movimento radial \'e obtida fazendo $\dd s^2=0$ no elemento de linha (\ref{Schw}) e considerando $\dd\theta=0$ e $\dd\varphi=0$. Resulta} 

\bea                                                       \label{luz}%30
c\,\dd t=\epsilon_\gamma\frac{\dd r}{1-\rho/r}\,, 
\eea

\ppln{estante $\epsilon_\gamma=+1$ se signalo movi^gas en {\it pozitiva} direkto de $r$\,, kaj $\epsilon_\gamma=-1$ kontra^ue. Ni integras (\ref{luz}) por la unua signalo,}
\pprn{sendo $\epsilon_\gamma=+1$ se o sinal se move na dire\c c\~ao {\it crescente} de $r$\,, e $\epsilon_\gamma=-1$ no caso contr\'ario. Integramos a (\ref{luz}) para o primeiro sinal,}  

\bea                                                    \label{tempo1}%31
c\int_{t_1}^{t_2}\dd t=\epsilon_\gamma\int_{a}^b\frac{\dd r}{1-\rho/r}\,,  
\eea

\ppln{kaj anka^u por la dua signalo,}
\pprn{e tamb\'em para o segundo sinal,} 

\bea                                                    \label{tempo2}%32
c\int_{t_1+\dd_at}^{t_2+\dd_bt}\dd t=\epsilon_\gamma\int_{a+\dd_ar}^b\frac{\dd r}{1-\rho/r}\,.  
\eea

\ppln{Subtrahante (\ref{tempo1}) de (\ref{tempo2}) fari^gas}
\pprn{Subtraindo a (\ref{tempo1}) da (\ref{tempo2}) ocorre}  

\bea                                                  \label{diftempo}%33
c(\dd _bt-\dd_at) = \epsilon_\gamma\left(\int_{a+\dd_ar}^b-\int_{a}^b\right) \frac{\dd r}{1-\rho/r}\equiv-\epsilon_\gamma\int_{a}^{a+\dd _ar} \frac{\dd r}{1-\rho/r}\,, 
\eea

\ppln{tio estas,}
\pprn{isto \'e,} 

\bea                                                   \label{diftemp}%34
\dd_bt=\dd_at-\frac{\epsilon_\gamma}{c}\frac{\dd_ar}{1-\rho/a}\,.
\eea 

\ppln{Ni povas skribi $\dd_ar$ kiel funkcio de $\dd_at$ uzante $v$ difinata en (\ref{vel}) kun $\dd\varphi/\dd t=0$:}
\pprn{Podemos escrever $\dd_ar$ como fun\c{c}\~ao de $\dd_at$ usando a $v$ definida na (\ref{vel}) com $\dd\varphi/\dd t=0$:} 

\bea                                                        \label{va}%35
v_a=\epsilon_a\frac{1}{1-\rho/a}\frac{\dd_ar}{\dd_at}\,, 
\eea

\ppln{estante $\epsilon_a=+1$ se la fonto movi^gas en la pozitiva direkto de $r$ en momento de eligo, kaj $\epsilon_a=-1$ kontra^ue, por ke $v_a\geq0$. Portante $\dd_ar$ de (\ref{va}) al (\ref{diftemp}), ^ci tiu ekvacio fari^gas}
\pprn{sendo $\epsilon_a=+1$ se a fonte se movimenta no sentido crescente de $r$ no momento da emiss\~ao, e $\epsilon_a=-1$ no caso contr\'ario, para que $v_a\geq0$. Levando $\dd_ar$ da (\ref{va}) \`a (\ref{diftemp}), esta \'ultima equa\c c\~ao se torna}  

\bea                                                       \label{dbt}%36
\frac{\dd_at}{\dd_bt}=\frac{1}{(1+\epsilon\,v_a/c)}\,, \hspace{3mm} \epsilon:=-\epsilon_a\epsilon_\gamma\,. 
\eea 

\ppln{Kuna analizo de $\epsilon_a$ kaj $\epsilon_\gamma$ montras, ke  $\epsilon=+1$ se la fonto fori^gas de radiusa pozicio $r=b$ en momento de eligo de signaloj, kaj $\epsilon=-1$ kontra^ue. Kunigante  (\ref{dtaua}), (\ref{dtaub}), kaj (\ref{dbt}), ni fine ricevas}
\pprn{Uma an\'alise conjunta de $\epsilon_a$ e $\epsilon_\gamma$ mostra que $\epsilon=+1$ se a fonte est\'a se afastando da posi\c c\~ao radial $r=b$ no momento da emiss\~ao dos sinais, e $\epsilon=-1$ no caso contr\'ario. Juntando (\ref{dtaua}), (\ref{dtaub}), e (\ref{dbt}), n\'os finalmente recebemos}   

\bea                                                    \label{Dfinal}%37
D=\sqrt{\frac{1-\rho/a}{1-\rho/b}}\frac{1-\epsilon v_a/c}{\sqrt{1-{v_a}^2/c^2}}\,. %\,, \hspace{3mm} {\rm ou} \hspace{3mm} 
%D= \sqrt{\frac{1-\rho/a}{1-\rho/b}} \sqrt{\frac{1-\epsilon v_a/c}{1+\epsilon v_a/c}}\,. 
\eea 

\ppl{Ni vidas en (\ref{Dfinal}), ke se gravito ne estas, t.e., se $\rho=0$, okazas}
\ppr{Vemos na (\ref{Dfinal}) que na aus\^encia de gravita\c c\~ao, isto \'e, se $\rho=0$, ocorre} 

\bea                                                       \label{Dsr}%38
D_{sr}=\sqrt{\frac{1-\epsilon V/c}{1+\epsilon V/c}}\,, \hspace{3mm} V:=|\dd_ar/\dd t|\,; 
\eea 

\ppln{^ci tiu estas la rezulto proponita por special-relativeco.}
\pprn{este \'e o resultado proposto pela relatividade especial.} 

\ppl{Ekvacio (\ref{Dfinal}) anta^udiras, ke kuna ago de $v_a\neq0$ kaj gravito povas malaperigi Doppleran efikon; fakte, se $a>b$ do povas okazi $D=1$ se $\epsilon=+1$, tio estas, se la fonto foriras de pozicio $r=b$ en momento de eligo. Kontra^ue, se $a<b$ do povas okazi $D=1$ se $\epsilon=-1$, tio estas, se la fonto proksimi^gas al pozicio $r=b$ en momento de eligo.}
\ppr{A (\ref{Dfinal}) prediz que uma a\c c\~ao conjunta de $v_a\neq0$ e gravita\c c\~ao pode fazer desaparecer efeito Doppler; com efeito, se $a>b$ ent\~ao pode ocorrer $D=1$ se $\epsilon=+1$, isto \'e, se a fonte estiver se afastando da posi\c c\~ao $r=b$ no momento da emiss\~ao. Opostamente, se $a<b$ ent\~ao pode ocorrer $D=1$ se $\epsilon=-1$, isto \'e, se a fonte estiver se aproximando da posi\c c\~ao $r=b$ no momento da emiss\~ao.} 

\ppl{Ni nun kalkulas la Doppleran faktoron se la fonto estas tuje anta^u $b$, a^u tuje post. Por tio ni faras $a=b$ en (\ref{Dfinal}) kaj ricevas}
\ppr{Vamos agora calcular o fator Doppler se a fonte estiver imediatamente antes de $b$\,, ou imediatamente depois. Para isso fazemos $a=b$ na (\ref{Dfinal}) e recebemos}  

\bea                                                      \label{a22}%39
D=\sqrt{\frac{1-\epsilon v_a/c}{1+\epsilon v_a/c}}\,. 
\eea 

\ppln{Ru^gi^go ($D<1$) okazas se la fonto foriras de ricevanto  ($\epsilon=+1$) kaj violi^go ($D>1$) okazas se ^gi alproksimi^gas ($\epsilon=-1$). Ekvacio (\ref{a22}) similas al (\ref{Dsr}) de special-relativeco se $v_a$ estas la rapido $V$ de fonto la^u la special-relativeco.}
\pprn{Ocorre desvio para vermelho ($D<1$) se a fonte estiver se afastando do receptor ($\epsilon=+1$) e desvio para violeta se ela estiver se aproximando ($\epsilon=-1$). A (\ref{a22}) \'e id\^entica \`a (\ref{Dsr}) da relatividade especial se $v_a$ for a velocidade $V$ da fonte segundo a relatividade especial.}

\ppl{Se la fonto movi^gas geodeze kaj radiuse, kaj se $a>b$, do  (\ref{Dfinal}) estas skribebla multe pli kompakte. Tiuokaze, nomu $v_b$ la rapido de fonto kiam ^gi pasas preter $b$, kaj nomu $v_a$ la rapido kiam ^gi pasas preter $a$. ^Car la konstanteco de $E$ en (\ref{Egeral}) implicas}  
\ppr{Se a fonte tiver movimento geod\'etico radial, e se $a>b$, ent\~ao a (\ref{Dfinal}) pode ser escrita de forma muito mais compacta. Neste caso, chame $v_b$ a velocidade da fonte quando ela passa por $b$, e chame $v_a$ a velocidade dela quando passa por $a$. Como a const\^ancia de $E$ na (\ref{Egeral}) implica}   

\bea                                               \label{konstanteco}%40 
\frac{\sqrt{1-\rho/a}}{\sqrt{1-{v_a}^2/c^2}}=\frac{\sqrt{1-\rho/b}}{\sqrt{1-{v_b}^2/c^2}}\,, 
\eea

\ppln{tial (\ref{Dfinal}) simpli^gas al} 
\pprn{ent\~ao a (\ref{Dfinal}) simplifica para}

\bea                                                \label{Dfinalgeod}%41
D=\gamma_b(1-\epsilon v_a/c)\,, \hspace{3mm} \gamma_b:=1/\sqrt{1-{v_b}^2/c^2}\,. 
\eea 

\ppln{Ekvacio (\ref{Dfinalgeod}) validas anka^u se $a<b$, kondi\-^ce ke $\sqrt{(1-\rho/a)/(1-\rho/b)}\geq\sqrt{1-{v_a}^2/c^2}$\,; ^ci tiu estas  necesa kondi^co por ke fonto pasu tra la radiusa pozicio $r=b$.}
\pprn{A equa\c c\~ao (\ref{Dfinalgeod}) tamb\'em vale se $a<b$\,, contanto que $\sqrt{(1-\rho/a)/(1-\rho/b)}\geq\sqrt{1-{v_a}^2/c^2}$\,; esta \'e a condi\c c\~ao necess\'aria para que a fonte passe pela posi\c c\~ao radial $r=b$.}  

\ppl{Ni ^generaligas (\ref{Dfinal}) por okazo de anka^u \linebreak[1]observanto radiuse movi^ganta. Kalkulante kiel anta^ue ni ricevas}
\ppr{Vamos generalizar a (\ref{Dfinal}) para o caso de tamb\'em o observador estar em movimento radial. Calculando como anteriormente n\'os recebemos} 

\bea                                                   \label{DDfinal}%42
D=\sqrt{\frac{1-\rho/a}{1-\rho/b}}\sqrt{\frac{1-\epsilon v_a/c}{1+\epsilon v_a/c}}\sqrt{\frac{1-\epsilon'v_b/c}{1+\epsilon' v_b/c}}\,, 
\eea 

\ppln{estante $\epsilon'=+1$ se observanto foriras de radiusa pozicio $r=a$ en momento de recepto de signaloj, a^u $\epsilon'=-1$ kontra^ue, kaj estante $v_b$ la rapido~(\ref{vel}) de observanto en momento de recepto.}
\pprn{sendo $\epsilon'=+1$ se o receptor estiver se afastando da posi\c c\~ao radial $r=a$ no momento da recep\c c\~ao dos sinais, ou $\epsilon'=-1$ no caso contr\'ario, e sendo $v_b$ a velocidade~(\ref{vel}) do receptor no momento da recep\c c\~ao.} 

\ppsection{Apliko al Tero}{Aplica\c{c}\~ao \`a Terra}

\ppln{Ni aplikas nun iom da niaj teoriaj rezultoj al movado de satelitoj orbitante Teron. Unue, ni kalkulas radiuson de orbito de geodeze Tero-staranta satelito. ^Car Tera rotacirapido estas $\omega_T=(2\pi/86.400)\,rad/s$, kaj {\Sch a} radiuso de Tero estas $\rho_T=2Gm_T/c^2=9\times10^{-3}\,m$, tial el (\ref{wgeodc}) oni ricevas $r_{geod}=42\times10^6\,m$. Uzante tiun rezulton kaj (\ref{Drarb}) oni povas kalkuli la Doppleran efikon observatan en Tero ($r_T=6\times10^6\,m$) el geodeze Tero-staranta satelito: $D_{s\rightarrow T}\approx1 + 6\times10^{-10}$\, (violi^gon).}
% 8,87\,mm$, % 42.243\,km$. 
% Tiu valoro estas nur iomete pli granda ol la longo de Tera ekvatoro. 
% Tiu valoro estas $6,64$ fojoj la radiuso de Tero, $r_T=6.366\,km$\,. 
\pprn{Vamos agora aplicar alguns dos nossos resultados te\'oricos ao movimento de sat\'elites que orbitem a Terra. Primeiro n\'os calculamos o raio da \'orbita de um sat\'elite geod\'etico geoestacion\'ario. Como a velocidade angular da Terra \'e $\omega_T=(2\pi/86.400)\,rad/s$, e o raio de \Sch\ da Terra \'e $\rho_T=2Gm_T/c^2=9\times10^{-3}\,m$\,, da (\ref{wgeodc}) se recebe $r_{geod}=42\times10^6\,m$\,. Usando esse resultado e a (\ref{Drarb}) pode-se calcular o efeito Doppler observado na Terra a partir do sat\'elite geod\'etico geoestacion\'ario: $D_{s\rightarrow T}\approx1 + 6\times10^{-10}$ (desvio para o violeta).}

\ppl{Nun ni kalkulas radiuson de orbito de ne-geodeza satelito kun la sama rotacirapido de Tero kaj sen Dopplera efiko kiam vidata de Tero. Uzante la superajn valorojn en (\ref{r3}) oni ricevas $r_2=151\times 10^6\,m$. Atentu, ke esperinde tio estas pli granda ol radiuso de geodeze Tero-staranta orbito; anka^u rimarku, ke tio estas pli eta ol la distanco Tero-Luno.} 
% $r_2=150.810\,km$\,. 
\ppr{Agora n\'os calculamos o raio da \'orbita de um sat\'elite n\~ao-geod\'etico com mesma velocidade angular que a da Terra, e sem efeito Doppler quando visto da Terra. Usando os valores acima na (\ref{r3}) recebe-se  $r_2=151\times 10^6\,m$\,. Note que, como esperado, isso \'e maior que o raio da \'orbita geod\'etica geoestacion\'aria; note tamb\'em que isso \'e menor que a dist\^ancia Terra-Lua.}

\ppl{Du fre^sdataj artikoloj enhavas kelkajn el niaj rezultoj. Fakte, iom da cirklaj movadoj en nia sekcio 7 estas anka^u priskribitaj en~\cite{AbraBaj}, kaj la radiusaj movadoj en nia sekcio 9 estas studitaj en~\cite{GronBraeck}. Same kiel ni, tiuj artikoloj komparas $\Delta\tau$, sed ^ci tie ni plu analizas Dopplerajn efikojn.}
\ppr{Dois artigos recentes cont\^em alguns de nossos resultados. Com efeito, os movimentos circulares na nossa se\c c\~ao 7 est\~ao tamb\'em descritos em~\cite{AbraBaj}, e os movimentos radiais da nossa se\c c\~ao 9 s\~ao estudados em ~\cite{GronBraeck}. Do mesmo modo que n\'os, aqueles artigos comparam $\Delta\tau$, mas aqui n\'os analisamos tamb\'em efeitos Doppler.} 

\ppl{En estonta artikolo~\cite{reltemp2} ni studos plurajn difinojn de
intertempo en arbitraj spacotempoj en ^general-relativeco, kaj
generaligas iom da rezultoj de ^ci tiu artikolo. Estas interesa anka^u detale studi malkune la kontribuojn de special-relativeco kaj de ^general-relativeco al rezultoj de ^ci tiu artikolo.} 
\ppr{Em um futuro artigo~\cite{reltemp2} n\'os vamos estudar v\'arias defini\c c\~oes de intertempo em espa\c cotempos arbitr\'arios na relatividade geral, e vamos generalizar alguns dos resultados deste artigo. \'E interessante tamb\'em estudar detalhadamente as contribui\c c\~oes em separado da relatividade especial e da relatividade geral aos resultados deste artigo.} 

%\newpage
\ppdu{\vspace{2em}
\ppl{\section*{~}\vspace{-1em}} \nopagebreak
\ppR{\section*{Refer\^encias}\vspace{-1em}} \ppp \nopagebreak
\vspace{-1.9em}}
\selectlanguage{esperanto}

\ppdu{\end{Parallel}}
\end{document}